\documentclass[aps,prd,twocolumn,showpacs,floats,floatfix,letterpaper,nofootinbib,superscriptaddress,]{revtex4}

\usepackage{amssymb,amsmath,latexsym,mathrsfs}
\usepackage{graphicx}
\usepackage{epsfig}
\usepackage{multirow}
\usepackage{array}
\usepackage{color}

\newcommand{\bea}{\begin{eqnarray}}
\newcommand{\eea}{\end{eqnarray}}

\begin{document}


\title{Beyond six parameters: extending $\Lambda$CDM}

\author{Eleonora Di Valentino}
\affiliation{Institut d'Astrophysique de Paris (UMR7095: CNRS \& UPMC- Sorbonne Universities), F-75014, Paris, France}
\author{Alessandro Melchiorri}
\affiliation{Physics Department and INFN, Universit\`a di Roma ``La Sapienza'', Ple Aldo Moro 2, 00185, Rome, Italy}
\author{Joseph Silk}
\affiliation{Institut d'Astrophysique de Paris (UMR7095: CNRS \& UPMC- Sorbonne Universities), F-75014, Paris, France}
\affiliation{AIM-Paris-Saclay, CEA/DSM/IRFU, CNRS, Univ. Paris VII, F-91191 Gif-sur-Yvette, France}
\affiliation{Department of Physics and Astronomy, The Johns Hopkins University Homewood Campus, Baltimore, MD 21218, USA}
\affiliation{BIPAC, Department of Physics, University of Oxford, Keble Road, Oxford
OX1 3RH, UK}

\begin{abstract}
Cosmological constraints are usually derived under the assumption of a $6$ parameters $\Lambda$-CDM theoretical
framework or simple one-parameter extensions. In this paper we present, for the first time, cosmological 
constraints in a significantly extended scenario, varying up to $12$ cosmological parameters simultaneously,
including the sum of neutrino masses, the neutrino effective number, the dark energy equation of state, 
the gravitational waves background and the running of the spectral index of primordial perturbations. 
Using the latest Planck 2015 data release (with polarization)  
we found no significant indication for extensions to the standard $\Lambda$-CDM
scenario, with the notable exception of the angular power spectrum lensing amplitude, $A_{\rm lens}$ that is
larger than the expected value at more than two standard deviations even when combining the Planck data
with BAO and supernovae type Ia external datasets. In our extended cosmological framework, we find that a 
combined Planck+BAO analysis constrains the value of the r.m.s. density fluctuation parameter to $\sigma_8=0.781_{-0.063}^{+0.065}$  at $95 \%$ c.l., helping to relieve the possible tensions with the
CFHTlenS cosmic shear survey.  We also find a lower value for the reionization optical depth $\tau=0.058_{-0.043}^{+0.040}$ at $95$ \% c.l. respect to the one derived under the assumption of $\Lambda$-CDM. The scalar spectral index $n_S$ is now compatible
with a Harrison-Zeldovich spectrum to within $2.5$ standard deviations. Combining the Planck dataset with the HST
prior on the Hubble constant provides a value for the equation of state $w < -1$ at more than two standard deviations
while the neutrino effective number is fully compatible with the expectations of the standard three neutrino framework.
\end{abstract}


\maketitle

\section{Introduction}

In the past twenty years, measurements of the Cosmic Microwave Background (CMB, hereafter) 
anisotropy angular power spectrum have  witnessed one of the most impressive technological 
advances in experimental physics. Following the first detection of CMB temperature anisotropies at large angular scales 
 by the COBE satellite in $1992$ \cite{COBE}, the angular power spectrum has been
measured with increasing precision by balloon-borne experiments such as BOOMERanG \cite{boom},
MAXIMA \cite{maxima}  and by ground-based experiments as DASI \cite{dasi},
 showing the unambiguous presence of a "first peak" and subsequent 
oscillations in the angular power spectrum at intermediate angular scales ($\theta \sim 0.2^o$).
The spectacular measurements obtained by the WMAP satellite mission \cite{WMAP}  have not only confirmed
the presence of these acoustic oscillations but also provided the first precise measurement of
the cross temperature-polarization angular spectrum and the first constraints on the epoch of reionization.
The very small-scale part of the angular temperature power spectrum, and especially the damping tail,  has been accurately determined
by experiments as ACT \cite{ACT} and SPT \cite{spt}. 
This impressive progress in the measurement of the CMB anisotropies temperature angular spectrum has
culminated with the cosmic-variance limited measurements of the Planck experiment 
that has now also provided exquisite results on the polarization and cross temperature-polarization spectra.

Despite this impressive progress on the experimental side, it is interesting to note that the
constraints on cosmological parameters are still presented (as in the latest Planck 2015 data release,
\cite{planckparams2015})  under the assumption of a simple $\Lambda$-CDM model, based on the variation
of just $6$ cosmological parameters. While this model still provides a good fit to the data, it
is the same model used, for example, in the analysis of the BOOMERanG $1998$ data
(see \cite{boom}), i.e. more than fifteen years ago. 
While this "minimal" approach is justified by the good fit to the data that the $\Lambda$-CDM provides 
we believe that it does not do adequate justice to the high quality of the most recent datasets.
In light of the new precise data, some of the assumptions or simplifications 
made in the $6$ parameters approach are indeed not anymore 
fully justified. For example, fixing the total neutrino mass to zero or to some small value is completely arbitrary since
we know that neutrinos must have masses and that current cosmological datasets are sensitive to
variations in the absolute neutrino mass scale of order $\sim 100$ meV.  
At the same time, considering that a cosmological constant offers 
difficulties in any theoretical interpretation, it seems reasonable to incorporate in the analysis a possible  dynamical dark energy component. This  is certainly plausible
(and even preferred if one wants to address the "Why Now ?" problem), and indeed fixing the dark energy 
equation of state to $-1$  is not favoured by any theoretical argument.
Most  inflationary models  predict a sizable contribution of gravitational waves.
Given the progress made in the search for B-mode polarization, especially by the recent combined BICEP2+Planck 
analysis \cite{BKP}, it is an opportune moment to allow any such 
contribution to be directly constrained by the data, without assuming a null contribution as in the 6-parameter model.  
A similar argument can be made for the running of the scalar spectral index
$dn_s/dlnk$. Moreover, the neutrino effective number, $N_{eff}$ could be easily different from the standard
expected value of $3.046$. Even assuming the standard three neutrino framework, non-standard decoupling,
 inflationary reheating, dark matter decay and many other physical processes could alter its value.
Finally, the Planck 2015 release still hints for an anomalous value for the lensing amplitude $A_{\rm lens}$
\cite{cala}. While this parameter is purely phenomenological, one should clearly consider it and check if the cosmology
obtained is consistent with other datasets.

Clearly, one should be careful in neglecting these extensions following a motivation based simply on 
an Occam razor argument. Indeed having more parameters does not necessary means having a more 
complicated theory. For example, forcing the dark energy component to a cosmological constant means
accepting an extreme fine tuning and a physical theory probably vastly more "complicated" as those
arising from a multiverse scenario.
 
The goal of this {\it Letter} is to constrain cosmological parameters in this extended parameter space.

\section{Method}

As discussed in the introduction, besides the six parameters of the "standard" $\Lambda$-CDM model, i.e.  
the Hubble constant $H_0$, the baryon $\Omega_bh^2$ and 
CDM energy densities $\Omega_ch^2$, the primordial amplitude and spectral index 
of scalar perturbations $A_s$ and $n_s$ (at pivot scale $k_0=0.05 hMpc^{-1}$), 
and the reionization optical depth $\tau$, we also consider variations in  
$6$ additional parameters: the total mass for the $3$ standard neutrinos, $\Sigma m_\nu$, the 
dark energy equation of state $w$ assumed  constant with redshift, the 
tensor/scalar ratio of amplitude $r$ at pivot scale $k_0=0.05 hMpc^{-1}$,
the running of the scalar spectral index $dn_s/dlnk$,  at pivot scale $k_0=0.05 hMpc^{-1}$, 
the amplitude of the lensing signal in the CMB angular spectra, $A_{\rm lens}$ as defined in \cite{cala},
the effective number of relativistic neutrinos, $N_{\rm eff}$.  
In what follows, we refer to this model as $e$CDM (extended Cold Dark Matter).

We let all these parameters vary freely simultaneously in a range of external, conservative, priors
listed in Table \ref{priors}.

\begin{table}
\begin{center}
\begin{tabular}{c|c}
Parameter                    & Prior\\
\hline
$\Omega_{\rm b} h^2$         & $[0.005,0.1]$\\
$\Omega_{\rm cdm} h^2$       & $[0.001,0.99]$\\
$\Theta_{\rm s}$             & $[0.5,10]$\\
$\tau$                       & $[0.01,0.8]$\\
$n_s$                        & $[0.8, 1.1]$\\
$\log[10^{10}A_{s}]$         & $[2,4]$\\
$\sum m_\nu$ (eV)               & $[0,3]$\\
$w$ & [-3.5,0.5]\\
$\frac{dn_s}{dlnk}$ & [-0.5,0.5]\\
$r$ & [0,0.5]\\
$N_{\rm eff}$ & [0.05,10]\\
$A_{\rm lens}$ & [0,10]\\
\end{tabular}
\end{center}
\caption{External priors on the cosmological parameters assumed in this paper.}
\label{priors}
\end{table}

\begin{table*}[t]
\begin{center}
\scalebox{0.77}{\begin{tabular}{|c|c|c|c|c|c|c|c|c|c|c|c|c|}
\hline
Model & & & & & &  &  & & & & &\\
Dataset & $\Omega_{\rm b}h^2$ & $\Omega_{\rm c}h^2$ & $H_0$ [km/s/Mpc] & $\tau$ & $n_s$ & $\sigma_8$ & $\frac{dn_s}{dlnk}$ & $r$ & $w$ & $\Sigma m_{\nu} [eV]$ & $N_{\rm eff}$ & $A_{\rm lens}$\\
\hline\hline
$\Lambda $ CDM  &  &  &  &  &  &  &  & & & & &  \\
    Planck TT+LowP& $0.02222^{+0.00046}_{-0.00044}$ & $0.1198^{+0.0042}_{-0.0043}$  & $67.3^{+2.0}_{-1.8}$ & $0.077^{+0.038}_{-0.036}$ & $0.966^{+0.012}_{-0.012}$ & $0.829^{+0.028}_{-0.028}$
&-&-&-&-&-&-\\
\hline
$\Lambda $ CDM  &  &  &  &  &  &  &  & & & & &  \\
    Planck& $0.02226^{+0.00031}_{-0.00029}$ & $0.1198^{+0.0028}_{-0.0028}$  & $67.3^{+1.3}_{-1.3}$ & $0.079^{+0.034}_{-0.035}$ & $0.9646^{+0.0092}_{-0.0092}$  & $0.831^{+0.026}_{-0.026}$ 
&-&-&-&-&-&-\\
\hline
$\Lambda $ CDM  &  &  &  &  &  &  &  & & & & & \\
Planck+ BAO  & $0.02229^{+0.00028}_{-0.00027}$ & $0.1193^{+0.0021}_{-0.0020}$  & $67.52^{+0.93}_{-0.93}$ & $0.082^{+0.031}_{-0.032}$ & $0.9662^{+0.0078}_{-0.0079}$ & $0.832^{+0.025}_{-0.025}$ 
&-&-&-&-&-&-\\
\hline
$e$ CDM&  &  &  &  &  &  & & & & & & \\
Planck TT+LowP  & $0.0245^{+0.0024}_{-0.0022}$ & $0.127^{+0.017}_{-0.016}$  & $>43$ & $0.073^{+0.051}_{-0.051}$ & $1.06^{+0.10}_{-0.098}$ & $0.56^{+0.35}_{-0.27}$  & $-0.004^{+0.042}_{-0.041}$ & $<0.383$ & $-0.53^{+0.61}_{-0.96}$ & $<1.30$ & $4.66^{+2.3}_{-2.1}$ & $2.50^{+2.3}_{-1.7}$  \\
\hline
$e$ CDM&  &  &  &  &  &  & & & & & & \\
Planck  & $0.02239^{+0.00060}_{-0.00056}$ & $0.1186^{+0.0071}_{-0.0068}$  & $>51.2$ & $0.058^{+0.040}_{-0.043}$ & $0.967^{+0.025}_{-0.025}$ & $0.81^{+0.24}_{-0.26}$  & $-0.003^{+0.020}_{-0.019}$ & $<0.183$ & $-1.32^{+0.98}_{-0.85}$ & $<0.959$ & $3.08^{+0.57}_{-0.51}$ & $1.21^{+0.27}_{-0.24}$  \\
\hline
$e$ CDM&  &  &  &  &  &  & & & & & & \\
Planck+BAO  & $0.02251^{+0.00056}_{-0.00052}$ & $0.1185^{+0.0069}_{-0.0069}$  & $68.4^{+4.3}_{-4.1}$ & $0.058^{+0.041}_{-0.043}$ & $0.972^{+0.024}_{-0.024}$ & $0.781^{+0.065}_{-0.063}$ 
& $-0.004^{+0.018}_{-0.018}$ & $<0.187$ & $-1.04^{+0.20}_{-0.21}$ & $<0.534$ & $3.11^{+0.52}_{-0.48}$ & $1.20^{+0.19}_{-0.19}$  \\
\hline
$e$ CDM&  &  &  &  &  &  & & & & & & \\
Planck+lensing & $0.02214^{+0.00053}_{-0.00052}$ & $0.1176^{+0.0069}_{-0.0066}$  & $>54.5$ & $0.058^{+0.040}_{-0.043}$ & $0.959^{+0.024}_{-0.024}$ & $0.85^{+0.21}_{-0.24}$ 
& $-0.005^{+0.018}_{-0.018}$ & $<0.178$ & $-1.45^{+0.96}_{-0.83}$ & $<0.661$ & $2.93^{+0.51}_{-0.48}$ & $1.04^{+0.16}_{-0.15}$  \\
\hline
$e$ CDM&  &  &  &  &  &  & & & & & &  \\
Planck+HST    & $0.02239^{+0.00059}_{-0.00057}$ & $0.1187^{+0.0072}_{-0.0070}$  & $74.4^{+5.1}_{-5.1}$ & $0.057^{+0.040}_{-0.045}$ & $0.966^{+0.025}_{-0.025}$ & $0.81^{+0.10}_{-0.11}$ 
& $-0.003^{+0.020}_{-0.019}$ & $<0.186$ & $-1.32^{+0.29}_{-0.31}$ & $<0.957$ & $3.09^{+0.58}_{-0.55}$ & $1.18^{+0.19}_{-0.18}$  \\
\hline
$e$ CDM&  &  &  &  &  &  & & & & & & \\
Planck+JLA   & $0.02242^{+0.00058}_{-0.00056}$ & $0.1188^{+0.0071}_{-0.0067}$  & $67.4^{+4.4}_{-4.2}$ & $0.058^{+0.040}_{-0.043}$ & $0.968^{+0.025}_{-0.025}$ & $0.759^{+0.088}_{-0.089}$ 
& $-0.004^{+0.020}_{-0.019}$ & $<0.183$ & $-1.06^{+0.13}_{-0.14}$ & $<0.854$ & $3.10^{+0.57}_{-0.54}$ & $1.20^{+0.19}_{-0.17}$  \\
\hline
$e$ CDM&  &  &  &  &  &  & & & & & & \\
Planck+WL & $0.02251^{+0.00056}_{-0.00055}$ & $0.1188^{+0.0073}_{-0.0069}$  & $>54.2$ & $<0.0835$ & $0.972^{+0.024}_{-0.024}$ & $0.82^{+0.22}_{-0.25}$ 
& $0.000^{+0.020}_{-0.019}$ & $<0.197$ & $-1.41^{+0.98}_{-0.79}$ & $<0.974$ & $3.16^{+0.58}_{-0.56}$ & $1.24^{+0.23}_{-0.22}$  \\
\hline
$e$ CDM&  &  &  &  &  &  & & & & & & \\
Planck+BAO-RSD & $0.02253^{+0.00052}_{-0.00050}$ & $0.1184^{+0.0069}_{-0.0067}$  & $68.6^{+4.2}_{-3.9}$ & $0.056^{+0.038}_{-0.042}$ & $0.972^{+0.023}_{-0.023}$  & $0.774^{+0.055}_{-0.058}$ 
& $-0.004^{+0.018}_{-0.018}$ & $<0.188$ & $-1.05^{+0.17}_{-0.19}$ & $<0.626$ & $3.12^{+0.51}_{-0.48}$ & $1.22^{+0.18}_{-0.17}$  \\
\hline
$e$ CDM&  &  &  &  &  &  & & & & & & \\
Planck+BKP &   $0.02237^{+0.00057}_{-0.00056}$ & $0.1186^{+0.0072}_{-0.0069}$  & $>52.3$ & $0.058^{+0.039}_{-0.044}$ & $0.966^{+0.026}_{-0.026}$ & $0.81^{+0.23}_{-0.25}$ 
& $-0.003^{+0.019}_{-0.018}$ & $<0.101$ & $-1.31^{+0.96}_{-0.89}$ & $<0.876$ & $3.07^{+0.57}_{-0.55}$ & $1.20^{+0.24}_{-0.22}$  \\
\hline

\hline\hline
\end{tabular}}
\caption{Constraints at $95 \%$ c.l. on the cosmological parameters assuming the standard $6$-parameter $\Lambda$CDM model
and the extended, $12$-parameter, $e\Lambda$CDM model.}
\label{table2}
\end{center}
\end{table*}

\begin{figure}
\centering
\includegraphics[scale=0.7]{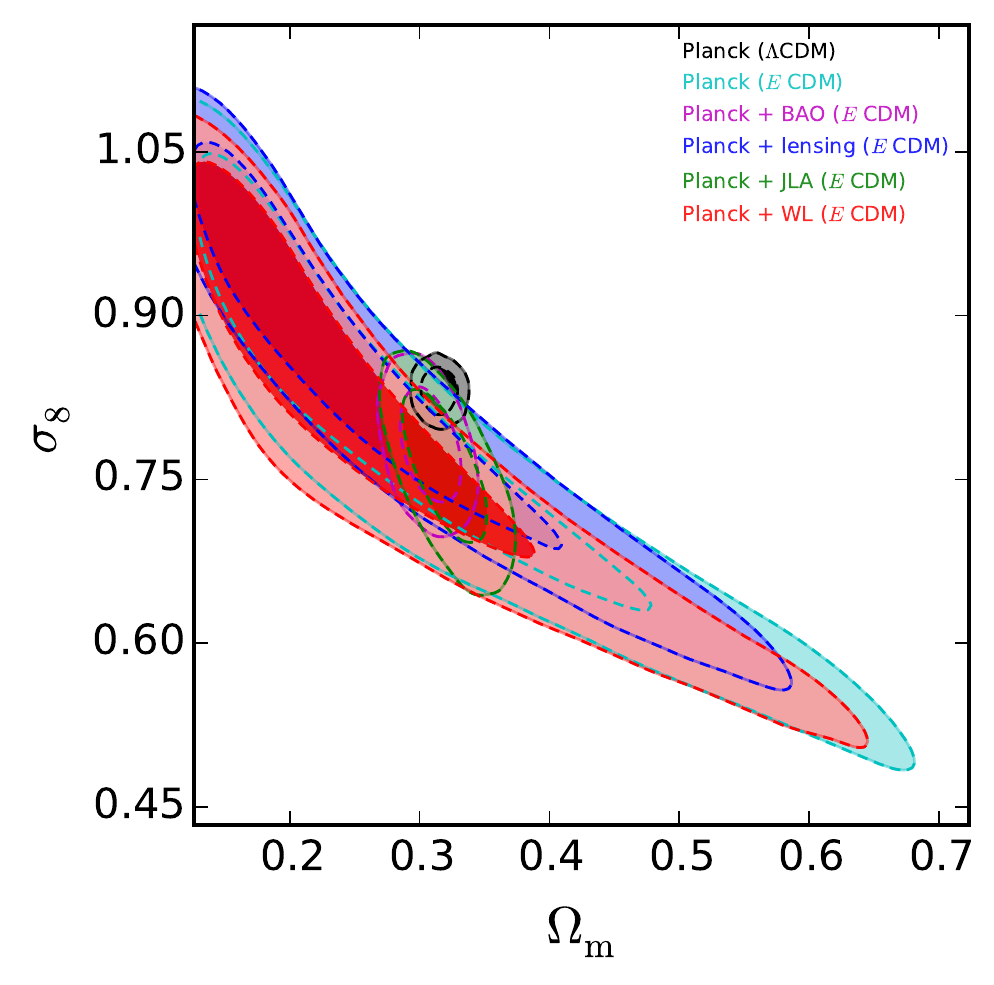}
\caption{Constraints at $68 \%$ and  $95 \%$ confidence levels on the $\sigma_8$ vs $\Omega_m$  plane under the assumption of 
$e\Lambda$CDM and different datasets. Black contours are the constraints under $\Lambda$CDM.}
\label{fig1}
\end{figure}

\begin{figure}
\centering
\includegraphics[scale=0.7]{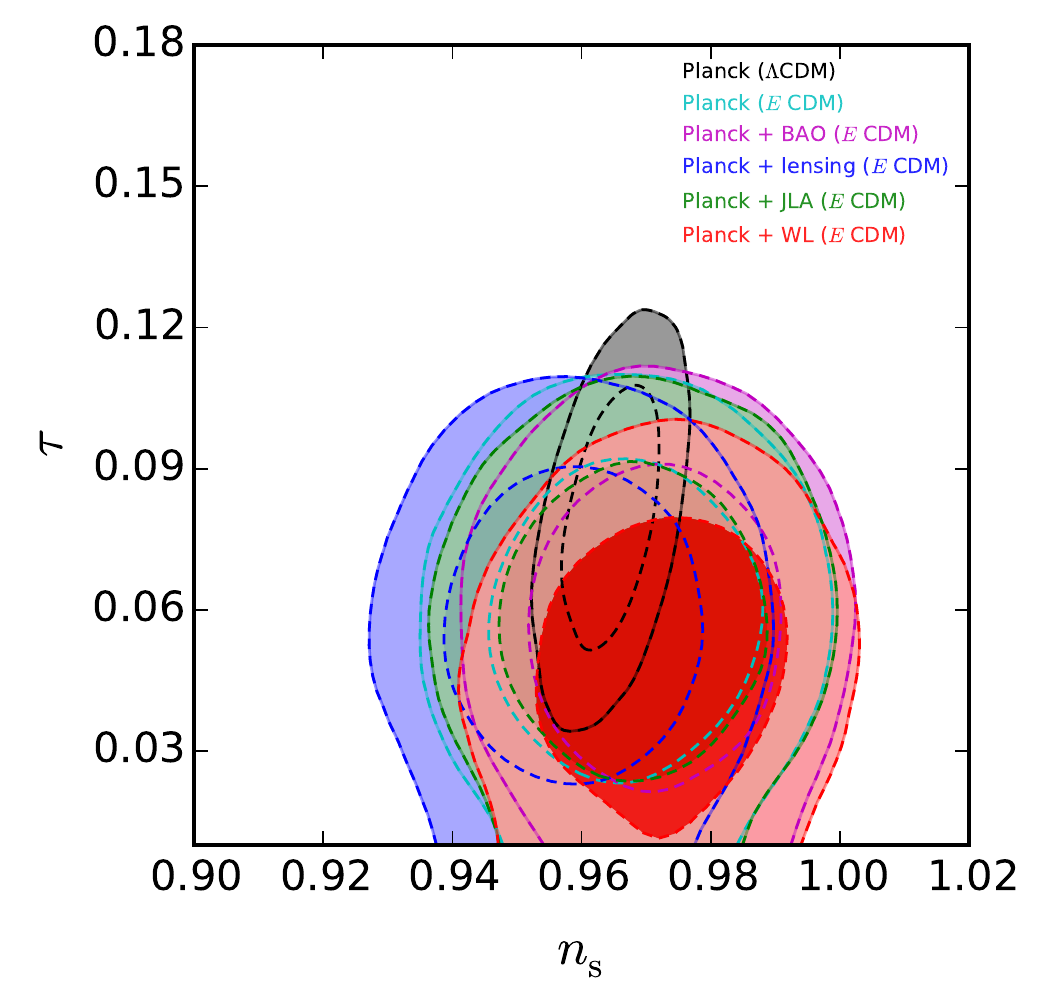}
\caption{Constraints at $68 \%$ and  $95 \%$ confidence levels on the $\tau$ vs $n_S$ plane under the assumption of $e\Lambda$CDM and different datasets. Black contours are the constraints under $\Lambda$CDM.}
\label{fig2}
\end{figure}

We produce constraints on these cosmological parameters by making use of several, recent, datasets.
Firstly, we use the full Planck 2015 release on temperature and polarization 
CMB angular power spectra. Since unresolved systematics may be present in the Planck 2015
polarization data at small angular scales (see discussion in \cite{plancklike2015}) we denote
as Planck TT+lowP the Planck temperature data plus the low angular scale polarization, while
with Planck the dataset that also includes the small-scale polarization data measured by Planck HFI.

 We also include  information on CMB lensing from Planck trispectrum detection (see \cite{plancklens2015}). We refer to this dataset
as {\it lensing}. We add  baryonic acoustic oscillation data from 6dFGS \cite{beutler2011}, SDSS-MGS \cite{ross2014}, 
BOSSLOWZ \cite{anderson2014} and CMASS-DR11 \cite{anderson2014} surveys as in \cite{planckparams2015}. 
We refer to this dataset as BAO. We impose a constraint on the Hubble constant from the Hubble Space Telescope \cite{HST} dataset.
Recently this constraint has been criticized in \cite{GHST} where a more conservative
value was suggested, a choice adopted in the recent Planck analysis
\cite{planckparams2015}. Since the HST prior has been shown to be in tension with the
Planck constraints on the Hubble constant obtained under $\Lambda$-CDM, we choose to use the less conservative
\cite{HST} determination in order to investigate its compatibility in our extended scenario. We refer to this dataset as HST.
We use luminosity distances of supernovae type IA from the "Joint Light-curve Analysis"
derived from the SNLS and SDSS catalogs \cite{JLA}. We refer to this dataset as JLA.
We  add weak lensing galaxy data from the CFHTlenS \cite{WL} survey with the priors and conservative cuts
to the data as described in \cite{planckparams2015}. We refer to this dataset as
WL. We consider redshift space distortions from \cite{RSD} with the prescription give in \cite{planckparams2015}. We refer to this dataset as RSD. Finally, we include upper limits on CMB polarization $B$ modes as recently placed by a common
analysis of Planck, BICEP2 and Keck Array data \cite{BKP}. We refer to this dataset
as BKP.

We use the publicly available Monte Carlo Markov Chain package \texttt{cosmomc} \cite{Lewis:2002ah} with a convergence diagnostic based on the Gelman and Rubin statistic. We use the July 2015 version which includes support for the Planck data release 2015 Likelihood \cite{plancklike2015} (see \url{http://cosmologist.info/cosmomc/}) and implements an efficient sampling using the fast/slow parameter decorrelations \cite{Lewis:2013hha}.
While in this paper we will focus the attention on cosmological parameters, we also vary foreground parameters 
using the same technique and parametrization 
described in \cite{plancklike2015} and \cite{planckparams2015}.

\section{Results}

The results of our analysis are reported in Table \ref{table2} where we also include, for comparison, the constraints
obtained assuming the standard, $6$ parameters in  $\Lambda$CDM.
The significant increase in the number of parameters produces, as expected, a relaxation in the
constraints on the $6$ $\Lambda$CDM parameters. 
Considering the allowable volume of the six-dimensional
$\Lambda$CDM parameter space to be proportional to the square root of the determinant of the $6\times6$ 
parameter covariance, we find that moving from $\Lambda$CDM to $e\Lambda$CDM expands this volume
by a factor $\sim 63000$ in case of the Planck dataset.
The parameters that are mostly affected are the 
Hubble constant and the r.m.s. amplitude of density fluctuations that are now practically undetermined from
Planck measuments alone and have significantly larger errors with respect to $\Lambda$CDM 
even when external datasets such as BAO are included. The main reason for this relaxation is the inclusion
in the analysis of the dark energy equation of state $w$, that introduces a geometrical degeneracy with the
matter density and the Hubble constant.  Moreover, marginalizing over the lensing amplitude
$A_{\rm lens}$ removes the lensing information in the CMB spectra that could potentially break 
this geometrical degeneracy. In this respect, it is interesting to note that a Planck+HST analysis
provides a value for the  equation of state $w$ less than the cosmological constant value $-1$ 
at more than $95 \%$ c.l.. As mentioned, the HST prior we adopt is not conservative. This result shows
that in case the HST measurement will be confirmed by future data, the only parameter between the extra $6$ we consider 
that can be varied to acccomodate this tension is the dark energy equation of state $w$.  Increasing the neutrino number does not seem to offer a viable solution as was on the contrary the case with the previous Planck 2013 release.
 
Constraints on the baryon and cold dark matter densities, the scalar spectral index $n_S$ and the
optical depth $\tau$  are also much weaker, mainly due to degeneracies between these parameters and $A_{\rm lens}$ 
and $N_{\rm eff}$.
Apart from the increase in the errors, it is interesting that parameters as $\sigma_8$ and the optical depth $\tau$
are shifted toward lower values respect to $\Lambda$CDM. These shifts are clear in Figures \ref{fig1} and \ref{fig2}
where we plot the $68 \%$ and $95 \%$ c.l. contour plots in the $\sigma_8$ vs $\Omega_m$ and $\tau$ vs $n_S$ planes, respectively. This is mainly due to the anomalous value of $A_{\rm lens}$ and persists when external datasets as BAO, JLA, WL and RSD are included. Looking at the results in Table \ref{table2}, the
value of $A_{\rm lens}$ is always different from the standard value at more than $95 \%$ c.l. when the Planck
CMB dataset is combined with external datasets, with the only notable exception of  the lensing information from the Planck trispectrum that pushes the value of $A_{\rm lens}$ back to agreement with unity. The nature of the Planck $A_{\rm lens}$ anomaly
could be different from the lensing determination but since it also persists  in our extended $e\Lambda$CDM scenario, 
this clearly deserves further investigation.
Moreover, $A_{\rm lens}$ is the only parameter that hints at a tension 
with standard $\Lambda$CDM. Again, by looking at Table \ref{table2}, apart from $A_{\rm lesn}$, 
we  see no evidence for "new physics": we  just have (weaker) upper limits on the neutrino mass, the running of
the spectral index is compatible with zero, the dark energy equation of state is compatible with $w=-1$
(expect when we use the HST prior), and the neutrino effective number  is remarkably close to the standard value $N_{\rm eff}=3.046$. It is  impressive that even in a $12$ parameter space, the neutrino effective  number is
still constrained with exquisite precision. This is mainly due to the inclusion of the Planck HFI small angular scale 
polarization data in the analysis. Removing this dataset, but keeping the low angular scale LFI polarization, 
we get a much weaker constraint from Planck+BAO of $N_{\rm eff}=4.35_{-1.6}^{+1.8}$ at $95 \%$ c.l..
The Planck+BAO constraint on neutrino mass of $\Sigma m_{nu} <0.534$ eV at $95 \%$ c.l. is significantly 
weaker with respect to the constraint  $\Sigma m_{nu} <0.174$ eV at $95 \%$ c.l.. obtained with the same dataset but assuming $\Lambda$CDM. The constraint on the tensor/scalar amplitude $r$ is about a factor two larger than in $\Lambda$CDM.
However, when the BKP dataset is included, the $95 \%$ c.l. upper limit of $r<0.108$ is recovered. This clearly
shows how a measurement of primordial B modes is crucial to constrain the tensor amplitude in a model-independent
way. The inclusion of the BKP dataset affects only the constraint on the tensor amplltude and leaves the other constraints
virtually unchanged.

\section{Conclusions}
\label{sec:conclusions}
  
In this {\it Letter} we have presented, for the first time, constraints on cosmological parameters in the framework of
an "extended" cold dark matter model ($e\Lambda$CDM) that is based on $12$ parameters instead of the usual $6$ 
assumed in the $\Lambda$CDM model.
In this extension some of the parameters usually well constrained under
 $\Lambda$CDM such as the Hubble constant and the amplitude of matter density fluctuations $\sigma_8$
are now unconstrained by CMB observations. Combining the CMB data with several other datasets 
reveals no statistically significant evidence for any tensions. More specifically, 
we have found no evidence for "new physics" beyond the standard $\Lambda$CDM
model,  i.e. there is no island of parameters in our extended theoretical framework that
could be preferred to the standard $\Lambda$CDM territory.
However $e\Lambda$ CDM prefers a lower value of $\sigma_8$ relative to that
obtained for 6-parameter $\Lambda$CDM but still requires
 a slightly  anomalous value of $A_{\rm lens} >1$.
The lower value of $\sigma_8$ in $e\Lambda$CDM brings the Planck data
in more agreement with the results of the CFHTlenS survey \cite{mcrann}. 
This result  motivates further studies that could explain the physical
nature of the $A_{\rm lens}>1$ anomaly.

The tension between the Planck and HST values of the Hubble parameter, in the
$e\Lambda$CDM scenario is solved by a value of the dark energy equation of state
$w<-1$ while the neutrino effective number remains compatible with the standard value
of $3.04$. 

Of course, the number of parameters can be further extended by considering, for example,
non-zero  curvature, isocurvature primordial perturbations,  features in the primordial spectrum, 
a varying (with redshift) dark energy equation of state, non-standard Big Bang Nucleosynthesis and 
a change in the primordial helium abundance $Y_p$, and so on.
Further extensions however may be premature because of degeneracies. For example, most effects of varying curvature are degenerate with a variation in $w$. CDM isocurvature modes have a 
spectrum similar to a GW background and the $A_{\rm lens}$ parameter could
account for undetected features in the angular spectrum. Moreover, a change in $N_{\rm eff}$ could
account for a change in $Y_p$.

We find it impressive that despite the increase in the
number of the parameters, some of the constraints on  key parameters
are relaxed but not significantly altered.
The cold dark matter ansatz remains robust, 
the baryon density is compatible with BBN predictions,
and the neutrino effective number is compatible
with standard expectations. The excellent quality of the new data
motivates our exploration beyond  the overexploited territory of $\Lambda$CDM towards
new and uncharted frontiers. 

JS and EdV acknowledge support by ERC project 267117 (DARK) hosted by UPMC, and JS for support at JHU by National Science Foundation grant OIA-1124403 and by the Templeton Foundation. EdV has been supported in part by the Institute Lagrange de Paris. AM acknowledge support by the research grant Theoretical Astroparticle Physics number 2012CPPYP7 under the program
PRIN 2012 funded by MIUR and by TASP, iniziativa specifica INFN.

\end{document}